# Application of retardation-modulation polarimetry in studies of nanocomposite materials

Andriy V. Kityk
Faculty of Electrical Engineering
Czestochowa University of Technology
Czestochowa, Poland
andriy.kityk@univie.ac.at

Patrick Huber
Institute of Materials Physics and Technology
Hamburg University of Technology
Hamburg, Germany
patrick.huber@tuhh.de

Anatoliy Andrushchak
Institute of Telecommunications,
Radio Electronics and Electronic Technics
Lviv National Polytechnic University
Lviv, Ukraine
anat@polynet.lviv.ua

Przemysław Kula[1], Wiktor Piecek[2]
Faculty of Advanced Technologies and Chemistry
Military University of Technology
Warszawa, Poland
[1]przemyslaw.kula@wat.edu.pl
[2]wiktor.piecek@wat.edu.pl

***Abstract*—We demonstrate an application of retardation-modulation polarimetry in studies of nanocomposite materials. Molecular ordering is explored on both nonchiral and chiral liquid crystals (LCs) in the bulk state and embedded into parallel-arrays of cylindrical channels of alumina or silica membranes of different channel sizes (12-42 nm). Two arms polarimetry serves for simultaneous measurements of the birefringence retardation and optical activity characterizing, respectively, orientational molecular ordering and chiral structuring inside nanochannels.**

## I. INTRODUCTION

The properties of nanoporous materials continue to be an interesting problem for both basic science and technological reasons. Recent trends in electronics and optoelectronics shift many of applications to nanoscale level where nanocomposites appear to be in a priority position. Up to date relevant studies have been widely presented mainly by nanocomposites based on liquid crystals (LCs), among which nematics [1-3], discotics [4,5] or smectics [6-8] embedded into nanoporous substrates (matrices), were in focus of numerous experimental investigations [1-10] and/or computer simulations [11]. The molecular ordering of non-chiral LCs in tabular cylindrical nanopores is considerably affected by interfacial interactions, quenched disorder and geometrical confinement having crucial influence on both static and dynamic properties of relevant nanocomposite materials. Prominent examples are paranematic (PN) ordering in the interface region or considerably different rates of dipolar relaxation in the core and interface regions of the pore filling demonstrated in recent optical [2,3] and dielectric studies [9,10]. Chiral LCs may form, in addition, confined helical structures which depending on anchoring condition may be presented by a single- or double-twist configurations [12-14]. In a number of cases the relaxation dynamics of helical structures under nanoconfinement, as e.g. in confined ferroelectric LCs [12], becomes several orders of magnitude faster than that of the bulk which opens up prospects for high-speed optoelectronic applications.

In the present work we demonstrate retardation-modulation polarimetry technique and its application to study of the LCs based nanocomposites. Examples of nonchiral and chiral nematics will be considered. However, two arms polarimeter has more universal its applicability. It may be applied also in studies of other types of nanocomposite materials, particularly those which are combined of solid nanocrystals embedded into tabular pores of inorganic or organic nanoporous matrices.

## II. EXPERIMENTAL RESULTS AND DISCUSSION

The modulation-retardation polarimeter is depicted in Fig.1. Two arms serve here for simultaneous measurements of the birefringence retardation, $\Delta$, and optical activity (rotation), $\Psi$. In contrast to similar solutions [15], the polarimetry setup employs only one photoelastic modulator, PEM. The modulated lights intensities are detected by photodetectors, PD, and subsequently analysed by the two pairs of lock-in amplifiers that measure amplitudes of the first ($I_{\Omega}$) and second ($I_{2\Omega}$) harmonics. The acquired data are transferred via GPIB to PC for their saving and further processing. The measured optical retardation and rotation are determined using equations inserted in Fig.1. The effective coefficient, $k$, is determined in a prior calibration procedure.

We report here optical polarimetry experiments on nonchiral LC 2,3,2'-trifluoro-4-pentyl-4"-propyl-p-terphenyl (abbreviated hereafter as 5FPFFPP3) and chiral ester LC S-(+)-4-(2-methylbutyl)phenyl-4-decyloxybenzoate (CE-6), in the bulk state and embedded into parallel-arrays of cylindrical channels of alumina or silica membranes of different channel sizes. The molecular structures of 5FPFFPP3 and *CE*-6 LCs along with the phase diagrams in their bulk state are presented in Fig.2.

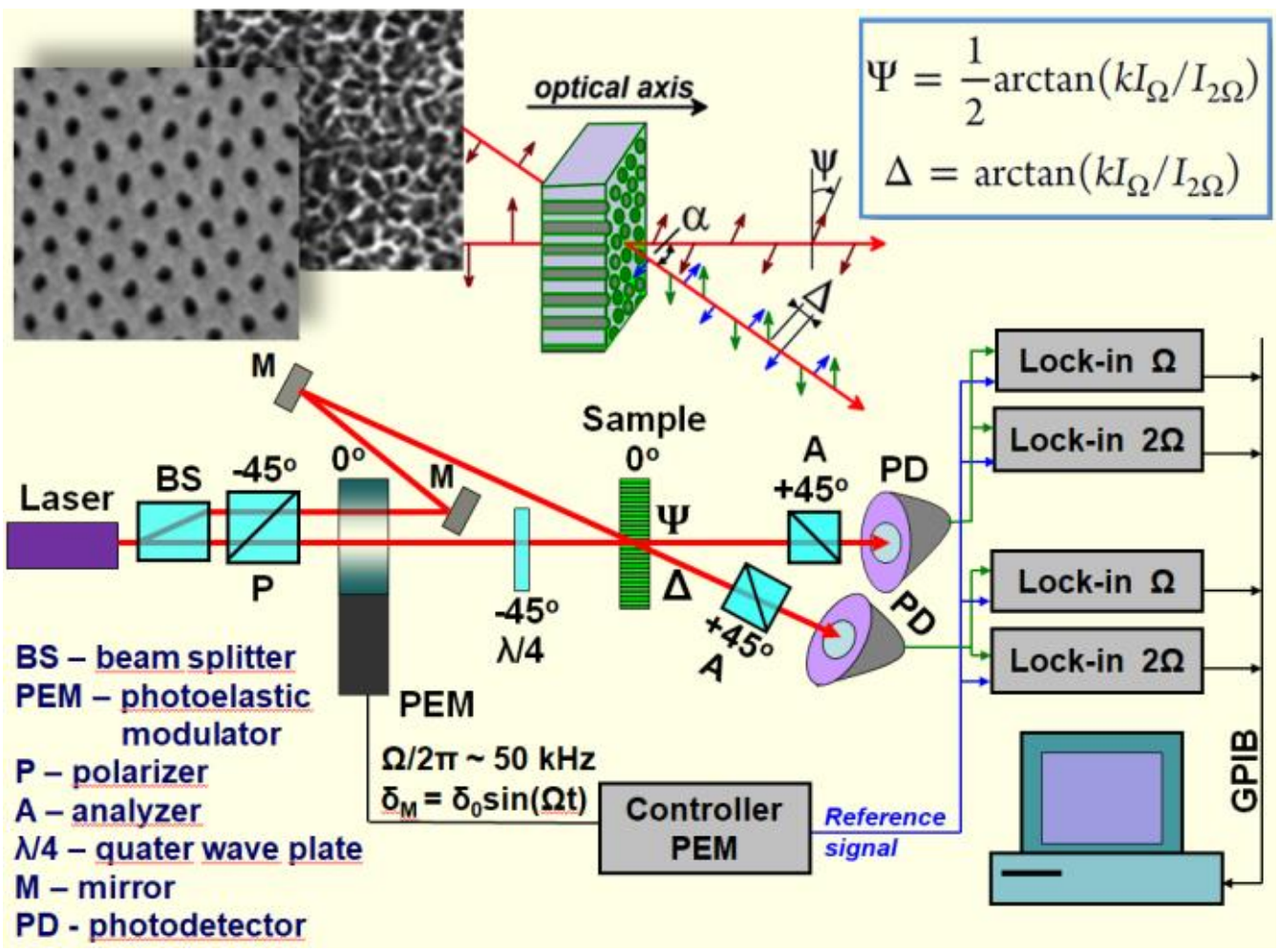

Fig.1. Two arms retardation-modulation polarimeter.

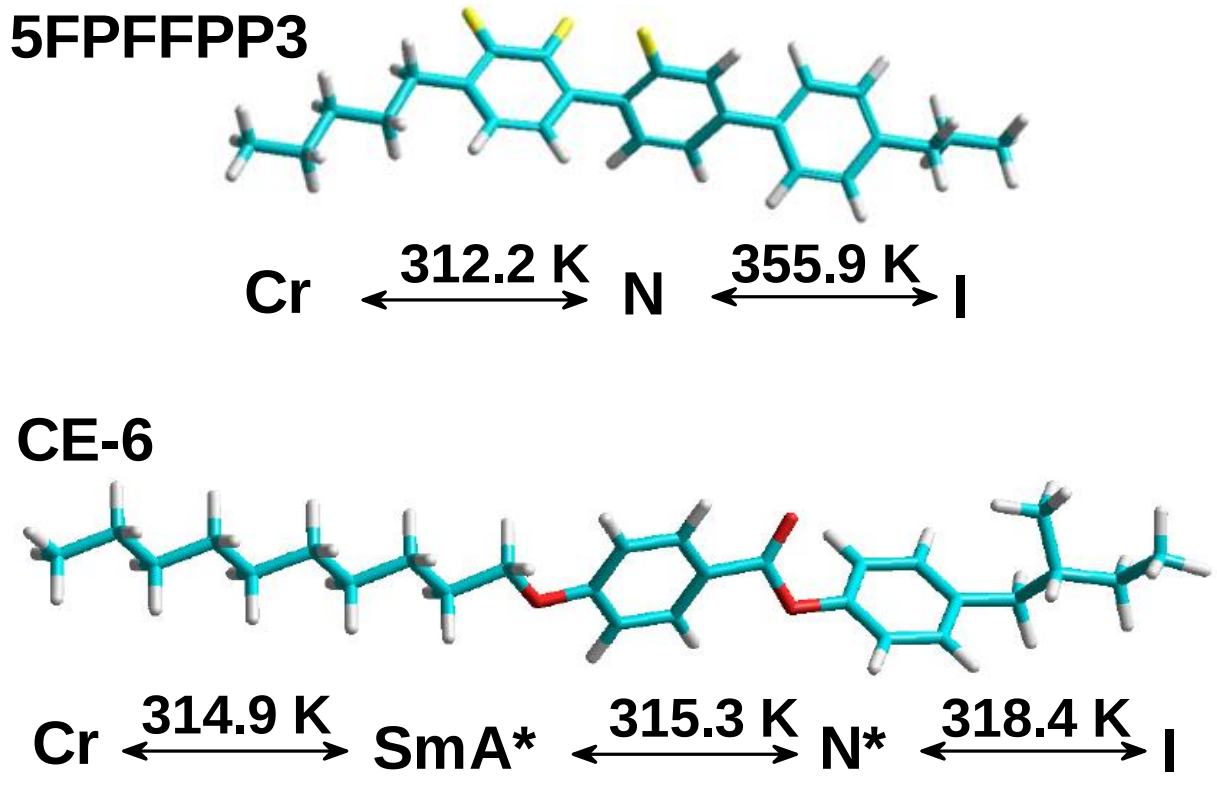

Fig.2. Schematic sketch of the nonchiral (5FPFFPP3) and chiral cholesteric (CE-6) LCs along with relevant LC phases and phase transition temperatures in the bulk state. I - isotropic phase, N – nematic phase, N* - chiral nematic phase, SmA* - chiral smectic A phase, Cr – crystalline phase.

In optical polarimetry experiments with nanoporous materials straight cylindrical pore geometry has a number of advantages comparing e.g. to nanoporous materials with random pore networks. In relevant composites preferable orientational ordering of the guest molecules results in an excess birefringence, or associated with it the excess retardation, which can be easily measured in a tilted sample geometry (see Fig.1). Fig.3 presents temperature dependences of the specific optical retardation, $R=\Delta/d$, where $d$ is the sample thickness, and the specific retardation normalized by the porosity $P$, $R^*= \Delta/(Pd)$, of nonchiral LC 5FPFFPP3 measured, respectively, in the bulk state [panel (a)] and embedded into parallel-arrays of cylindrical channels of alumina or silica membranes of different pore sizes [panel (b)]. For rodlike molecules the degree of orientational molecular ordering can be described by the order para-meter $Q = 0.5\cdot\langle\cos^2\varphi - 1\rangle$, where $\varphi$ is the angle between the long axis of particular molecules and the direction of preferred orientation (director). An excess optical retardation, $R^+$ ($R^{*+}$), induced by orientational molecular ordering, is proportional to the effective (averaged) order parameter $Q^*$ and for this reason may be used for its characterization. In the bulk state *R(T)*- dependence exhibits features being evident for the first order phase transformation from Isotropic(I)-to-Nematic (N) phase. At cooling relevant transition is characterized by a jump-like rising of the optical retardation whereas any pretransitional effects are evidently absent in the bulk isotropic phase, see Fig.3(a).

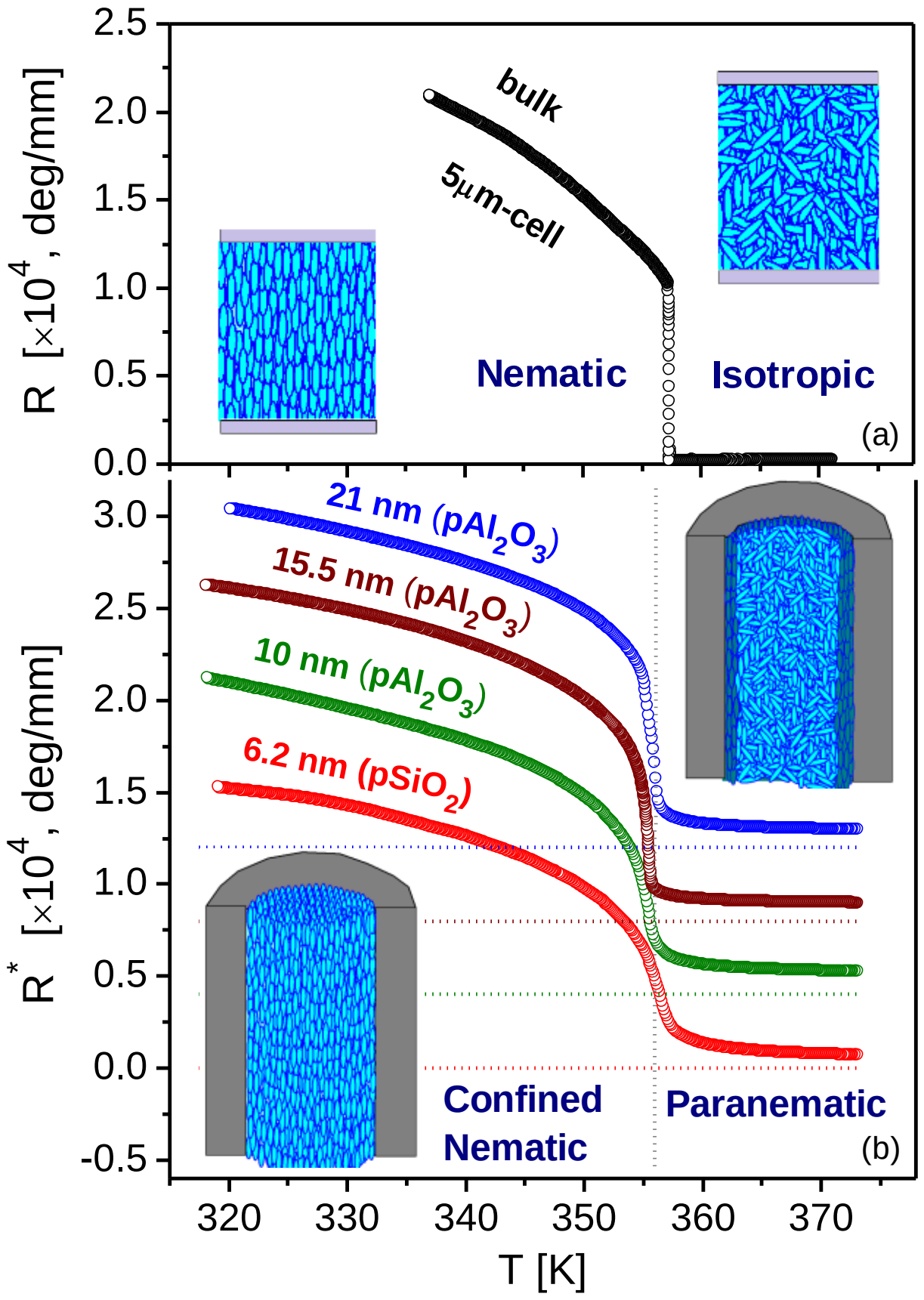

Fig.3. Temperature dependences of the specific optical retardation, *R*, and the porosity normalized specific optical retardation, *R**, of nonchiral LC 5FPFFPP3 measured, respectively, in the bulk state (a) and embedded into parallel-arrays of cylindrical channels of alumina or silica membranes of different pore radii, *r* (b), see labeled. Incident angle, α = 36°. Inserts show sketches of molecular ordering in the isotropic (paranematic) and nematic phases both in the bulk (a) and confined states (b).

The nanoconfined LC 5FPFFPP3 exhibits a substantially different behaviour, see Fig. 3(b). It is especially evident for the LC embedded into nanochannels of small pore sizes (pore radii $r \leq 10$ nm). Upon cooling $R^*$ increases continuously indicating on a preferable alignment of rodlike molecules parallel to the channel axes. Presumably native silica or alumina pore walls render here tangential anchoring resulting in positive birefringence, see inset in Fig.3(b). Moreover, interfacial interactions render also a continuous transition with residual birefringence extending far above I-N transition temperature. Accordingly, it leads to so-called paranematic (PN) LC state being a characteristic feature of confined nematic LCs [1-3]. Such behaviour, on the other hand, is less prominent for the LC confined into nanochannels of larger sizes, i.e. for channel radii $r \geq 15$ nm. Relevant changes of $R^*$ in the vicinity of PN-N transition appear to be more steeper in that case remembering much bulk behaviour.

Observed behaviour may be interpreted by a phenomenological Landau–de Gennes model within the approach suggested by Kutnjak, Kralj, Lahajnar, and Zumer (KKLZ model) [1,6]. The dimensionless free energy density describing spatially confined states in a cylindrical geometry and tangential anchoring conditions reads in the KKLZ model as

$$f = tq^2 - 2q^3 + q^4 - q\sigma + \kappa q^2 \ , \qquad (1)$$

where $q = Q/Q(T_{IN})$ and $t = (T\text{-}T^*)/(T_{IN}\text{-}T^*)$ are the reduced order parameter and the reduced temperature, respectively, $T^*$ is the effective temperature in the definition given in [1]. The last term in Eq. (1) describes so-called quenched disordering effects due to pore wall irregularities (surface-induced deformations) [2]. In the PN behaviour a key role belongs to the bilinear coupling term, $q\sigma$, where $\sigma \sim 1/r$ is so-called geometrical ordering field. A curved channel geometry enforces orientational molecular ordering along the channels, i.e. acts as an external field.

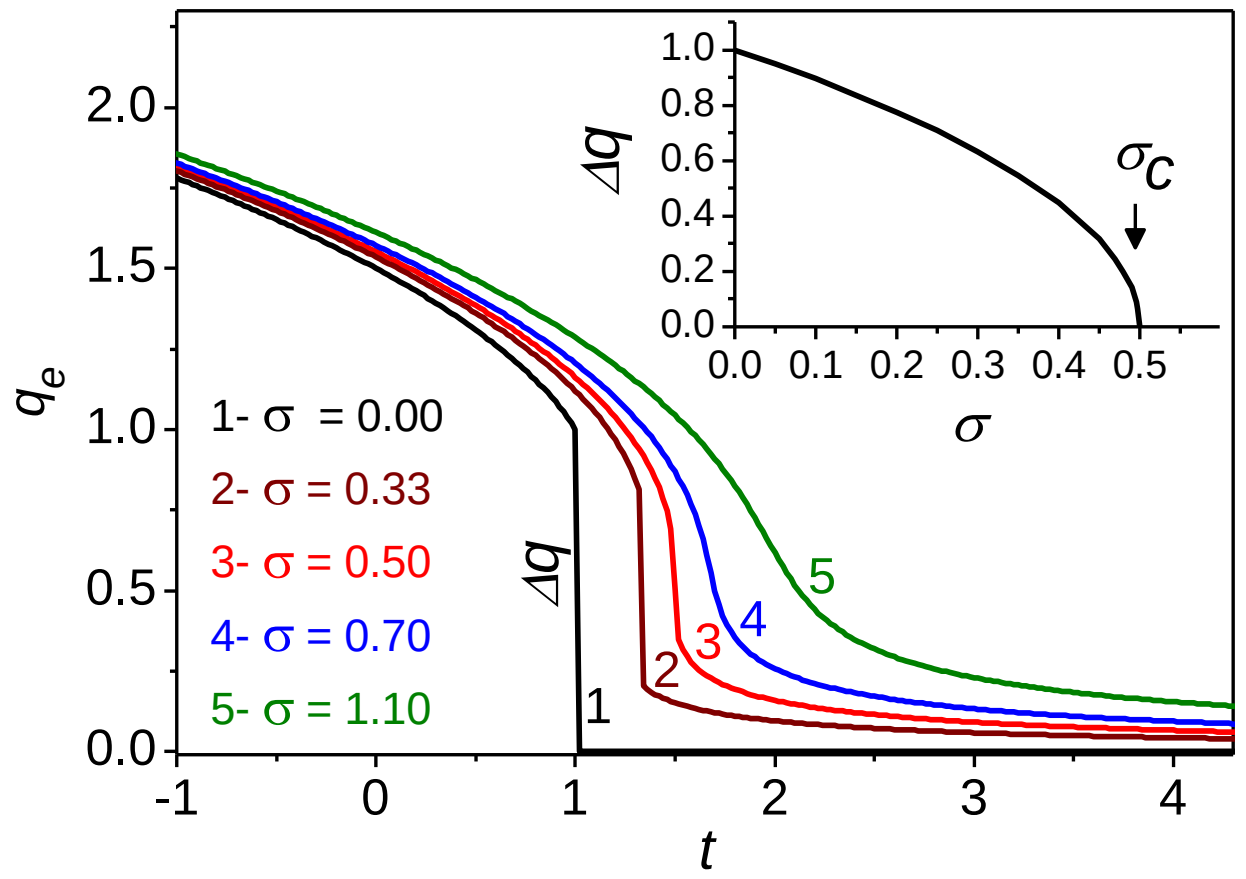

Fig.4. Equilibrium order parameter $q_e$ in the KKLZ model as a function of reduced temperature $t$ for selected effective surface fields, $\sigma$ (see labelled) in the absence of quenched disorder ($\kappa = 0$). Inset shows the order parameter jump at the I-N (P-N) transition as a function of the surface field $\sigma \sim 1/r$.

Minimization of the free energy $f$ yields the equilibrium order parameter $q_e$, which is shown in Fig. 4. Within KKLZ approach the PN-N transition is of the first order as long as the geometrical field $\sigma$ is less than its critical value equals 0.5. Above it the behaviour of the order parameter is continuous. At $\sigma = \sigma_C = 0.5$ one deals with so-called critical behaviour. Comparing the behaviour of the equilibrium order parameter, $q_e(t)$ (Fig.4) and the behaviour of measured normalized specific retardation depicted in Fig.3 one may conclude that at large channel sizes ($r \geq 15$ nm) we arrive in sub-critical regime. At small pore sizes ($r \leq 10$ nm) a supercritical behaviour is enforced by large geometrical ordering fields causing a continuous temperature evolution of the birefringence retardation in the vicinity of the PN-N transition. The critical channel radius value, $r_C$, corresponding to the critical ordering field $\sigma_C$, appears in the range between 10 and 15 nm. Relevant value is of the same order of magnitude as ones derived recently in a series of $n$CB LCs ($n$ = 5-7) [2,3].

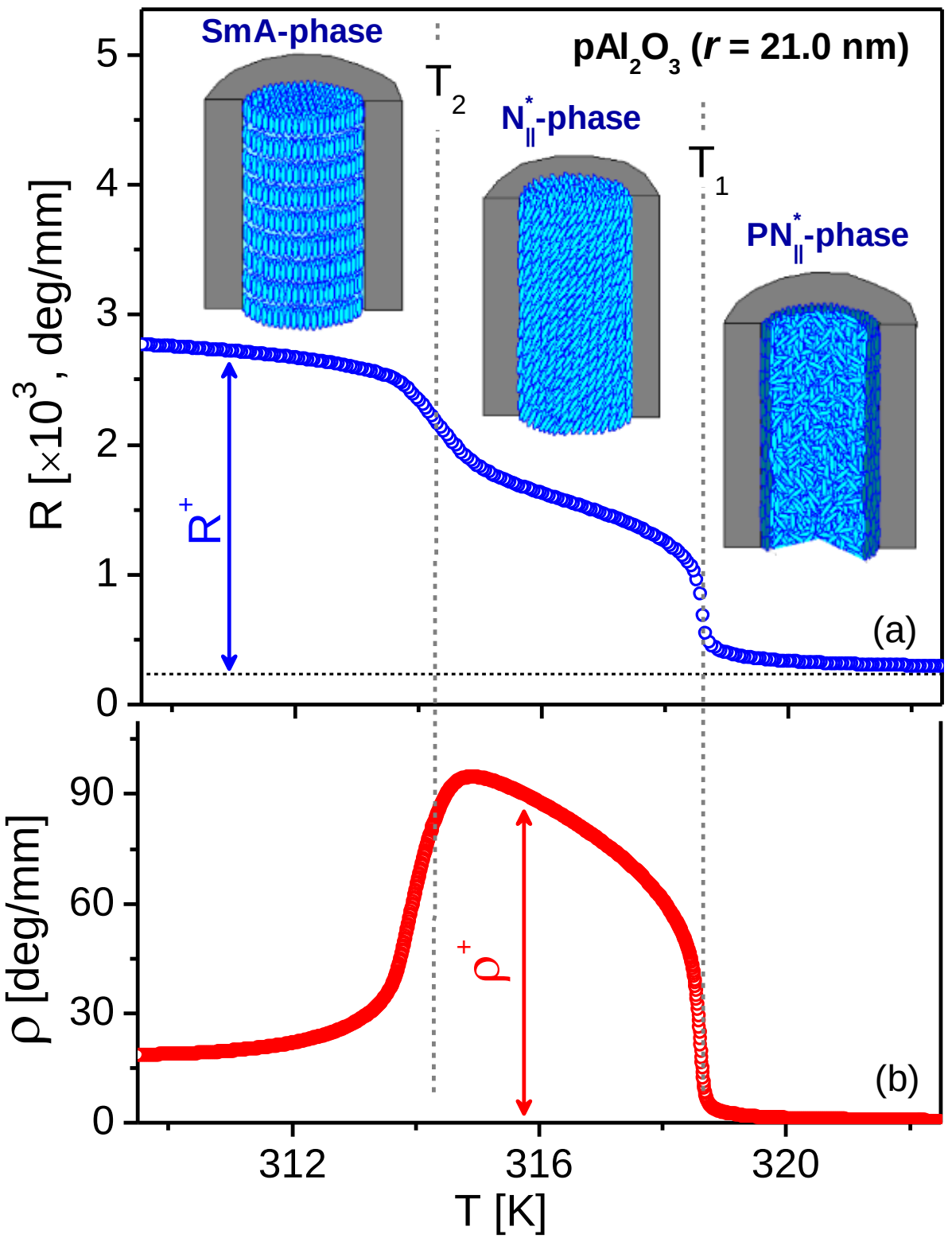

Fig.5. Specific optical retardation $R$ (a) and optical rotation $\rho$ (b) of CE-6 confined into cylindrical channels ($r$ = 21 nm) of the alumina membranes. Relevant excess values, $R^+$ and $\rho^+$, caused by cholesteric and smectic molecular ordering are shown. Tangential anchoring is provided by polymer (SE-130) wall coating. Inserts in section (a) show sketches of molecular ordering in the confined isotropic (I), chiral nematic (cholesteric, N*) and chiral smectic A (SmA*) phases.

The ordering of nonchiral nematics molecules inside the nanopores may be not uniform within the pore volume, however, in average it results in a linear birefringence only. Molecular chirality brings new features to the optical properties among which the circular birefringence, called also as optical activity (rotation of light polarization) represents another polarimetric characteristic being frequently measured for identification and/or characterization of helical structures in cholesterics or other chiral LCs such as e.g. ferroelectric LCs (SmC* phases) or twist bend nematic phases.

In present work we explore chiral ordering of the cholesteric LC CE-6 confined into cylindrical channels ($r$ = 21 nm) of the alumina membranes. In the bulk state cholesteric LC CE-6 exhibits a sequence of three phase transitions during cooling: it transforms from the isotropic (I) phase to the chiral nematic (N*) phase at 318.4 K, to the SmA* phase at 315.3 K and to the solid crystalline (Cr) phase at 314.9 K. However, the transition to the Cr-phase can be easily supercooled [13] in the confined geometry thus the ferroelectric SmA* phase can be observed in a somewhat broader temperature range, i.e. even down to 309 K.

Fig. 5 demonstrates specific optical retardation $R$ [panel (a)] and optical rotation $\rho$ [panel (b)] of CE-6 confined into cylindrical channels ($r$ = 21 nm). Tangential anchoring is provided here by SE-130 polymer coating of the channels walls. Relevant excess values, $R^+$ and $\rho^+$, caused by cholesteric and smectic molecular ordering, are shown. For such types of anchoring our optical polarimetry measurements indicate evidently on a sequence of two phase transitions. Above $T_1$ the nanocomposite shows small optical retardation and rotation characteristic for confined isotropic phases. Residual birefringence retardation indicates on a certain PN ordering originating from the interface region, i.e. it has the same origin as in confined nonchiral nematics. Below $T_1$ the retardation and optical activity both rise considerably. The excess birefringence is found to be positive. On the one hand, it means that we deal with molecular ordering characterizing by a preferable orientation of molecules being nearly parallel to the long channel axes. Anomalous rise of the optical activity below $T_1$, on the other hand, indicates on a formation here a long-range helical structure. It accompanies a local orientational order and apparently propagates along the channel axis. The resulting molecular ordering in the temperature range $T_1$-$T_2$ is sketched in insert of Fig.5(a). It may be represented by so-called double-twist structure ($N_{||}$*-phase [13]).

Step-like rising of the optical retardation in the vicinity of $T_2$ evidently indicates on a second phase transformation. It is accompanied by a considerable suppressing of the optical activity. Appropriate interpretation of such behavior, confirmed also in the X-ray measurements [13], is a formation of SmA*-phase with smectic layers oriented perpendicularly to the long pore axes, see sketch in Fig.5. Accordingly, below $T_2$ molecules order parallel to the long pore axis with no helical pitch what results in strengthening of the birefringence retardation and suppressing the optical activity.

One should notice that the arrangement of the chiral molecules CE-6 inside the nanochannels may be controlled by appropriate anchoring conditions. Applying, for instance, SE-1211 polymer coating a normal anchoring may be achieved. If this is the case, a simple single-twist structure ($N_{\perp}$*-phase [13]) becomes favorable below $T_1$ whereas SmA*-phase appears to be completely suppressed [13].

## III. Conclusions

We have demonstrated here an application of retardation-modulation polarimetry in studies of nanocomposite materials. Molecular ordering is explored on both nonchiral and chiral liquid crystals (LCs) in the bulk state and embedded into parallel-arrays of cylindrical channels of alumina and silica membranes of different channel sizes (12-40 nm). Two arms polarimetry serves for simultaneous measurements of the birefringence retardation and optical activity characterizing, respectively, orientational molecular ordering and chiral structuring inside nanochannels.

However, two arms polarimeter has more universal applicability. It may be efficient in a study of other types of nanocomposite materials, also those which are combined of solid nanocrystals embedded into tabular pores of inorganic or organic nanoporous matrices. Such materials may be deposited inside the nanochannels from relevant saturated solutions, employing cooling or evaporation methods. Among the candidates for guest materials one may consider here a series of water soluble proper ferroelectric crystals, like e.g. TGS [16,17], KDP [18,19] or GPI [20], efficient acoustooptic crystals with suitable elastic properties, as e.g. $Cs_2HgCl_4$ or $Cs_2HgBr_4$ [21,22], or efficient nonlinear optical crystal materials, such e.g. KDP or $Ag(NaNO_2)_2$ [23]. Several such nanocomposites, with incorporated ferroelectric crystals KDP [18,24,25] and TGS [26] have been recently synthesized. Using X-ray technique it was demonstrated that crystallographic orientation of this crystals along the channel axis is identical in different pores. Accordingly, such nanocomposite materials are expected to exhibit macroscopic anisotropy which can be characterized by the optical polarimetry technique as has been demonstrated in present study.

## Acknowledgment

This work is a part of a project that has received funding from the European Union’s Horizon 2020 research and innovation program under the Marie Skłodowska-Curie grant agreement No 778156.